\documentclass[reprint, amsmath, amssymb, aps]{revtex4-1}	
\usepackage{graphicx}
\usepackage{dcolumn}
\usepackage{bm}
\usepackage{mathrsfs}
\usepackage{tikz}
\usepackage{hyperref}
\input epsf.tex
\usepackage{amssymb}
\usepackage{amsmath}
\usepackage{amsthm}
\usepackage{amsopn}
\usepackage{ulem}
\usepackage{cancel}
\usepackage{color}
\usepackage{float}
\hypersetup{
    colorlinks=true,       
    linkcolor=blue,          
    citecolor=blue,        
    filecolor=magenta,      
    urlcolor=blue           
}

\begin{document}

\title{ Epistemic uncertainty from an averaged Hamilton-Jacobi formalism}

\author{M. J. Kazemi}
\email{kazemi.j.m@gmail.com}
\affiliation{ Department of Physics, Faculty of Science, Shahid Beheshti University, Tehran 19839, Iran}
\author{S. Y. Rokni}
\affiliation{Department of Physics, Faculty of Science, Qom University, Qom, Iran}


\begin{abstract}
In recent years, the non-relativistic quantum dynamics derived from three assumptions;  i) probability current conservation, ii) average energy conservation, and iii) an epistemic momentum uncertainty [A. Budiyono and D. Rohrlich, \href{https://doi.org/10.1038/s41467-017-01375-w}{Nat. Commun 8, 1306 (2017)}]. Here we show that, these assumptions can be derived from a natural extension of classical statistical mechanics.


\begin{description}
\item[PACS numbers]
03.65.Ta,  05.10.Gg
\end{description}

\end{abstract}
\maketitle

\textit{"... we cannot afford to neglect any possible point of view for looking at Quantum Mechanics and in particular its relation to Classical Mechanics. Any point of view which gives us any interesting feature and any novel idea should be closely examined to see whether they suggest any modification or any way of developing the theory along new lines"}\cite{Dirac 1951}.

\ \ \ \ \ \ \ \ \ \ \ \ \ \ \ \ \ \ \ \ \ \ \ \ \ \ \ \ \ \ \ \ \ \ \ \ \ \ \ \ \ \ P. A. M. Dirac (1951)

\section{Introduction} 
The difference between classical and quantum mechanics can be intuitively understood via uncertainty relations. In fact, the uncertainty relations lie at the core of quantum mechanics and can quantitatively determine other fundamental quantum features, such as quantum interference and quantum non-locality \cite{Uncertainty-nonlocality, Uncertainty-interference}. Remarkably, the non-relativistic Schr\"{o}dinger equation recently derived from three assumptions;  i) probability current conservation, ii) average energy conservation, and iii) an exact form of uncertainty relation \cite{Budiyono 2017, Hall 2002}.  This begs a question: Are there profound and natural principles determining this exact form \cite{Hanggi 2013, Budiyono 2020, Budiyono 2020-2}? Here we show that, all these three assumptions can be derived from a natural extension of classical statistical mechanics; an averaged version of classical Hamilton-Jacobi formalism. This make a new answer to an old-- but still interesting  \cite{Budiyono 2017, Schleich 2013, Hall 2014, Sebens 2015, Lindgren 2019}-- question: Can we reconstruct quantum mechanics as a clear modification of classical statistical mechanics?
Moreover, this axiomatic quantization method leads naturally to a reasonable formulation of mixed classical-quantum dynamics; i.e. Hall's theory \cite{Hall-book 2016}.

\section{Ontology and Notation}
 The proposed formalism can be applied to some different interpretations of quantum theory. Nonetheless, for clarity, here we only use of "Many Interacting World" (MIW) interpretation \cite{Hall 2014, Sebens 2015, Poirier 2012, Tipler 2014} (or very similar proposal which called "Real Ensemble" interpretation \cite{Real-Ensemble-1, Real-Ensemble-2}). In this interpretation, the wave-function dose not exist; instead there is a 
real ensemble of classical "worlds", and all quantum effects merely arise from a universal interaction between these worlds.
This interpretation leads to a reasonable solution for \textit{measurement problem} \cite{Sebens 2015}. Furthermore, it is supported by PBR theorem, which claims that the quantum state must be corresponds to something real in nature \cite{Real-Ensemble-1, PBR}. 

 Let us also fix our notation. Consider an ensemble of classical-like $n$-particle systems, \textit{worlds}, which state of any of them is determined by a point in phase-space, $(x,p)=(\bm{x}_1 ,..., \bm{x}_n; \bm{p}_1,..., \bm{p}_n)$.  The locale average of any classical "observable", $O(x,p)$, which is represented by $\bar O(x)$, is defined as
\begin{equation*}
\bar O(x) \equiv \frac{1}{\rho(x)}\int f(p,x) O(x,p) d^{3n}p,
\end{equation*}
where $f(x,p)$ is phase-space distribution and $\rho(x)\equiv\int f(x,p) dp$ is configuration-space distribution of worlds-ensemble. Moreover, the local variance, $\overline{O^2}-\bar{O}^2$, is  represented by $(\delta O )^2$, and similarly, associated global average and variance are defined as $\langle O\rangle \equiv N^{-1}\int \rho(x) \bar O(x)d^{3n}x$ and $(\Delta O)^2\equiv N^{-2}\int\rho(x)(\delta O)^2 d^{3n}x$, respectively. 
Note that, these global ensemble averages are comparable with expectation values of associated operator, $\langle \hat{O} \rangle_\psi$, in the standard Hilbert-space formalism. 

\section{Assumptions and Results}
 In fact we only use MIW ontology without considering any special form for inter-world interaction priorly. Instead, we assume that the time evolution of worlds ensemble, associated to a pure quantum state, is describable by an average version of classical Hamilton-Jacobi formalism: Explicitly, we assume that, there is a smooth function, $S(x,t)$, which is related to local average momentum as $\bar p_i =\partial_iS$, and satisfy flowing axioms.  
\\

\textbf{Axiom 1}\textit{- Average Hamilton-Jacobi equation} \cite{footnote-3-2}:
\begin{equation}\label{axiom1}
\int \rho (\frac{\partial S}{\partial t}+\bar{H})d^{3n}x=0.
\end{equation}

\textbf{Axiom 2}\textit{- Least average action} \cite{footnote-3-3}:
\begin{equation}\label{axiom2}
\delta \int \rho (\frac{\partial S}{\partial t}+\bar{H})d^{3n}x dt=0.
\end{equation}
where, $\bar{H}(x)$, is local average of classical Hamiltonian, $H=\sum_i \bm{p}_i^2/2m+V(x)$, which explicitly reads
$\bar{H}=\sum_i [(\nabla_i S)^2+(\delta \bm{p}_i)^2]/2m+V(x).$ It is worth noting that, the time evolution of an classical ensemble, which is describable by classical Hamilton-Jacobi formalism, satisfy these axioms \cite{footnote-3-3}. The main result of this work is that above axioms determine the time evolution of $\rho$ and $S$ up to some free parameters, which is include classical and quantum dynamics as special cases:  see following theorem.

\textbf{Theorem:} The axioms (\ref{axiom1}-\ref{axiom2}), beside the assumption that the local momentum variance is a  local function of probability density as $(\delta \bm{p}_i)^2=\mu_i(\rho, \partial\rho)$, leads to
\begin{equation}\label{mu}
(\delta \bm{p}_i)^2=\lambda_i^2 (\frac{\nabla_i\rho}{\rho})^2
\end{equation}
and following evolution equations, uniquely: 
\begin{equation}\label{Rt}
\frac{\partial \rho}{\partial t}=-\sum_i\nabla_i(\rho\frac{\nabla_i S}{m_i})
\end{equation}
\begin{equation}\label{St}
\frac{\partial S}{\partial t}=-\sum_i\frac{(\nabla_i S)^2}{2m_i}+\frac{\lambda_i^2}{2m_i}\frac{\nabla_i^2\sqrt{\rho}}{\sqrt{\rho}}-U(x)
\end{equation}
where $\lambda_i^2$s are \textit{non-negative} constants.

\textbf{Proof --} For simplicity, here we prove the theorem for one-particle case, however it is completely similar to the N-particle case. 
The axiom-\ref{axiom2} leads to following evolution equations: 
\begin{equation}\label{dPdt1}
	\frac{\partial \rho}{\partial t}=-\nabla(\rho\frac{\nabla S}{m})
\end{equation} 
\begin{equation}\label{dSdt1}
	\frac{\partial S}{\partial t}=-\frac{(\nabla S)^2}{2m}-U(x)-Q_0,
\end{equation}
in which $Q_0\equiv [\frac{\partial}{\partial \rho}-\partial_i \frac{\partial}{\partial \partial_i\rho}](\rho\mu).$
The equation (\ref{dSdt1}) beside the axiom-\ref{axiom1} leads to
\begin{equation}\label{intpH}
	\int \rho \bar H d^3x=\int \rho [\frac{(\nabla S)^2}{2m}+U(x)+Q_0] d^3x,
\end{equation} 
which, using the axiom-\ref{axiom2}, leads to
\begin{equation}\label{dSdt2}
	\frac{\partial S}{\partial t}=-\frac{(\nabla S)^2}{2m}-U(x)+Q_1.
\end{equation}
where $Q_1\equiv[\partial_{ij} \frac{\partial}{\partial \partial_{ij}\rho}-\partial_k \frac{\partial}{\partial \partial_k\rho}+ \frac{\partial}{\partial \rho}](\rho Q_0).$
The equation (\ref{dSdt2}) is consistent with equation (\ref{dSdt1}), if and only if 
\begin{equation}\label{Goz}
	Q_1=Q_0.
\end{equation}
Equation (\ref{Goz}) is a strong constrain on the local momentum variance. Moreover, since the local momentum variance, $(\delta \textbf{p})^2$, is a scaler under rotation transformation, it must be a function of the $\rho$ and $(\nabla\rho)^2$; i.e.  $(\delta \textbf{p})^2=\mu(\rho,\eta)$, where $\eta=(\nabla\rho)^2$. Therefore, the equation (\ref{Goz}), by tedious but straightforward calculations, leads to 
\begin{equation}
	\mu(\rho,\eta)=a\frac{\eta}{\rho^2}+\frac{b}{\rho}+c, 
\end{equation} 
where $a$, $b$ and $c$ are arbitrary constants.  It is easy to see that the equation (\ref{intpH}) leads to $b=0$, and also the constant $a$ must be a non-negative real number, because the $(\delta \textbf{p})^2$ has this property.  Hence, we  get $Q_0 =Q_1= \frac{-\lambda^2}{2m}\frac{\nabla^2\sqrt{\rho}}{\sqrt{\rho}}+c$, where $\lambda^2\equiv2ma$. Finally note that, in the equation (\ref{dSdt1}), the constant $c$ can be absorbed in external potential $U(x)$; In fact, it must be equal to zero, if we assume that the momentum  of particles ensemble associated  to a plane wave, $e^{i\textbf{p}_0.\textbf{x}/\hbar}$, be equal to $\textbf{p}_0$. $\blacksquare$

\section{Discussion}
Above theorem means the average H-J formalisem, axioms (\ref{axiom1}-\ref{axiom2}), determine the time evolution of $\rho$ and $S$ up to some unknown constants. Firstly, let us consider the case that all $\lambda_i$s are same; $\lambda_i=\lambda$. In this case, if we consider $\lambda=0$, the equations (\ref{St}) reduce to classical H-J equation, which, beside equation (\ref{Rt}), describe dynamics of a classical (non-interacting) ensemble. On the other hand, if we consider $\lambda=\hbar$, the equations (\ref{Rt}) and (\ref{St}) describe dynamics of a quantum system: although we do not consider any wave function fundamentally, one can define an effective wave function associated to worlds ensemble as $\psi=\sqrt{\rho}e^{iS/\hbar}$, which satisfies Schr\"odinger equation, 
\begin{equation}
i\hbar\frac{\partial\psi}{\partial t}=\sum_i\frac{\hbar^2}{2m_i}\nabla_i^2\psi+U(x)\psi.
\end{equation}
Moreover, the equation (\ref{mu}) ensures the uncertainty relation, $\Delta x_i \Delta p_i \geq \hbar$ \cite{Budiyono 2017, Budiyono 2019}. 

Another interesting case is that considering  $\lambda_i=\hbar$ for one part of the system and  $\lambda_i=0$ for the rest; This choice leads to configuration ensemble theory for mixed classical-quantum systems \cite{Hall-book 2016}. 
Finally note that, one can use other values, $0<\lambda<\hbar$, to making (at least a phenomenological model for) a smooth description of quantum-classical transition \cite{Richardson 2014, Mousavi-Miret 2018, Carlo 2012}. Nonetheless, in the rest of the discussion, we consider only pure quantum systems, i.e. $\lambda_i=\hbar$ for all components of the system. 

It worth noting that, the equation (\ref{mu}) can be derived from an "exact" form of uncertainty relation, $\delta x_i\delta p_i=\hbar$ \cite{Hall 2002}, and then using of the equation (\ref{mu}), the equations (\ref{Rt}) and (\ref{St}) can be derived from probability conservation and average classical energy conservation assumptions, respectively \cite{Budiyono 2017, Smolin 1986}. The advantage of our derivation is that, we priory do not assume a cornerstone of quantum theory, i.e uncertainty relation, however it is derived from our "average Hamilton-Jacobi" formalism. Moreover, the meaning of equation (\ref{mu}) in our interpretation  is  different with exact uncertainty approach \cite{Hall 2002}, Nelson's stochastic mechanics \cite{Nelson 1966, Guerra 1983, Smolin 1986}, and other similar formalism \cite{Lindgren 2019, Pena 1983, Santos 2006}: we do not consider any random motion or inherent fluctuations, and the equation (\ref{mu}) merely represents the statistical momentum width of worlds ensemble. 



\begin{thebibliography}{9999999999999}

\bibitem{Dirac 1951}
P. A. M, Dirac, The Relation of Classical to Quantum Mechanics, Proceedings of the second Canadian mathematical congress (University of Toronto), 10 (1951).











\bibitem{Uncertainty-nonlocality}
J. Oppenheim, and S. Wehner, 
\href{https://doi.org/10.1126/science.1192065}{Science,  \textbf{19},  (2010)}.

\bibitem{Uncertainty-interference} 
P. Coles, J. Kaniewski, and S. Wehner,  
\href{https://doi.org/10.1038/ncomms6814}{Nat Commun \textbf{5},(2014)}; 
S. D\"{u}rr and G. Rempe, 
\href{https://doi.org/10.1119/1.1285869}{Am. J. Phys, \textbf{68}, (2000)}; 
P. Busch, C. Shilladay,
\href{https://doi.org/10.1016/j.physrep.2006.09.001}{Physics Reports, \textbf{435}, (2006)}; 
P. Storey, S. Tan, M. Collett,  et al, 
\href{https://doi.org/10.1038/367626a0}{Nature 367, 626–628 (1994)}. 



\bibitem{Schleich 2013}
W. P. Schleich, D. M. Greenberger, D. H. Kobe, and M. O. Scully, 
\href{https://doi.org/10.1073/pnas.1302475110}{PNAS, \textbf{110}, (2013)}.
\bibitem{Budiyono 2017}
A. Budiyono, and D. Rohrlich,  
\href{https://doi.org/10.1038/s41467-017-01375-w}{Nat Commun, \textbf{8}, (2017)}.
\bibitem{Hall 2002}
 M. J. W. Hall and M. Reginatto, 
\href{https://iopscience.iop.org/article/10.1088/0305-4470/35/14/310}{J. Phys. A: Math. Gen. 35, 3289 (2002)};
J. W. M. Hall,
\href{https://link.aps.org/doi/10.1103/PhysRevA.64.052103}{Phys. Rev. A, \textbf{64}, (2001)}. 
\bibitem{Budiyono 2020}
A. Budiyono, 
\href{https://link.aps.org/doi/10.1103/PhysRevA.101.022102}{Phys. Rev. A, \textbf{101}, (2020)}.
\bibitem{Budiyono 2020-2}
A. Budiyono,  and H. K. Dipojono,
\href{https://link.aps.org/doi/10.1103/PhysRevA.102.012205}{Phys. Rev. A, \textbf{101}, (2020)}.

\bibitem{Hanggi 2013}
E. Hänggi, S. A. Wehner, 
\href{https://doi.org/10.1038/ncomms2665}{Nat Commun, \textbf{4}, (2013)}. 


\bibitem{Hall-book 2016}
M. J. W. Hall,  and M. Reginatto, 
\href{https://link.aps.org/doi/10.1103/PhysRevA.72.062109}{Phys. Rev. A, \textbf{72}, (2005)}; Ensembles on Configuration Space, \href{https://doi.org/10.1007/978-3-319-34166-8}{Springer International Publishing Switzerland, (2016)}.

\bibitem{Schleich 2013}
W. P. Schleich, D. M. Greenberger, D. H. Kobe, and M. O. Scully, 
\href{https://doi.org/10.1073/pnas.1302475110}{PNAS, \textbf{110}, (2013)}.
\bibitem{Hall 2014}
M. J. W. Hall,  D-A. Deckert,  and H. M. Wiseman, 
\href{https://link.aps.org/doi/10.1103/PhysRevX.4.041013}{Phys. Rev. X, \textbf{4}, (2014)}; B. Poirier, \href{https://link.aps.org/doi/10.1103/PhysRevX.4.040002}{Phys. Rev. X, \textbf{4}, (2014)}.
\bibitem{Sebens 2015}
C. T. Sebens, 
\href{https://doi.org/10.1086/680190}{Philosophy of Science, \textbf{82}, (2015)}; K.J. Bostr\"{o}m,  
\href{https://doi.org/10.1007/s40509-015-0046-6}{Quantum Stud.: Math. Found. \textbf{2}, (2015)}.
\bibitem{Lindgren 2019}
J. Lindgren, J. Liukkonen, 
\href{https://doi.org/10.1038/s41598-019-56357-3}{Sci Rep \textbf{9}, 19984 (2019)}.

\bibitem{Poirier 2012}
J. Schiff and B. Poirier, 
\href{https://doi.org/10.1063/1.3680558}{J. Chem. Phys, \textbf{136}, (2012)}.
\bibitem{Tipler 2014}
F. J. Tipler, %
\href{ https://doi.org/10.1073/pnas.1324238111}{PNAS, \textbf{111}, (2014)}.

\bibitem{Real-Ensemble-1}
L. Smolin,  
\href{https://doi.org/10.1007/s10701-012-9666-4}{Found Phys,  \textbf{42},  (2012)}; 
\bibitem{Real-Ensemble-2}
L. Smolin,  
\href{https://doi.org/10.1007/s10701-016-9994-x}{Found Phys 46, 736–758 (2016)}.  

\bibitem{PBR}
M. Pusey, J. Barrett,  and T. Rudolph,  
\href{https://doi.org/10.1038/nphys2309}{Nature Phys, \textbf{8}, (2012)}. 

\bibitem{footnote-3-2}
Perhaps, the meaning of "average" is more cleared, if we rewrite this axiom as $\int f(x,p) [\partial_t S+H(x,p)]dx dp=0$. 
\bibitem{footnote-3-3}
This variational principle is a direct generalization  of the variational formulation of classical H-J formalism, i.e. $\delta\int \rho [\partial_t S+H(x,\nabla S)]dx dt=0$, which leads to classical H-J  and continuity equations \cite{Holland-book 1993}. Similar generalized actions, also used in some other methods for derivation of Schroedinger equation 
\cite{Hall 2002, averaged-action}. Moreover, one can use of an "average Hamilton's equations" instead of this variational principle; i.e. $\partial_t \rho=\delta \langle H\rangle/\delta S$ and $\partial_t S=-\delta \langle H\rangle/\delta \rho$, where $\rho$ and $S$ are considered as canonical conjugate variables \cite{Hall-book 2016}. In addition, it is easy to see that, the equation-of-motions which finally we find from these axioms, (\ref{Rt}) and (\ref{St}), have clear physical meanings in our interpretation: probability and average classical energy conservation, $d\langle H\rangle/dt=0$, which together ensure conservation of the total classical energy of ensemble.
\bibitem{Holland-book 1993}
P. R. Holland, 
\href{https://doi.org/10.1017/CBO9780511622687}{Cambridge University Press, (1993)}.
\bibitem{averaged-action}
G. Gr\"{o}ssing, 
\href{https://doi.org/10.1016/j.physleta.2008.05.007}{Phys. Lett. A, \textbf{372}, (2008)};
M. Reginatto, 
\href{https://link.aps.org/doi/10.1103/PhysRevA.58.1775}{Phys. Rev. A, \textbf{58}, (1998)};
M. Atiq, M. Karamian,  and M. Golshani, 
\href{https://doi.org/10.1007/s10701-008-9260-y}{Found Phys, \textbf{39}, (2009)}. 


\bibitem{Budiyono 2019}
A. Budiyono, 
\href{https://link.aps.org/doi/10.1103/PhysRevA.100.062102}{Phys. Rev. A \textbf{100}, 062102 (2019)}.



\bibitem{Richardson 2014}
C. D. Richardson,  P. Schlagheck,  J.  Martin,  N. Vandewalle,  and T. Bastin, 
\href{https://link.aps.org/doi/10.1103/PhysRevA.89.032118}{Phys. Rev. A, \textbf{89}, (2014)}.

\bibitem{Mousavi-Miret 2018}
C-C. Chou, 
\href{https://doi.org/10.1016/j.aop.2016.06.001}{Annals of Physics, \textbf{371}, (2016)}; 
\href{https://doi.org/10.1016/j.aop.2018.04.017}{Annals of Physics, \textbf{393}, (2018)}; S. V. Mousavi and S. Miret-Artés, 
\href{https://doi.org/10.1016/j.aop.2018.04.009}{Annals of Physics, 393, (2018)}. 

\bibitem{Carlo 2012}
Gabriel G. Carlo, 
\href{https://link.aps.org/doi/10.1103/PhysRevLett.108.210605}{Phys. Rev. Lett, \textbf{108}, (2012)}; G. G. Carlo, L. Ermann, A. M. F. Rivas, M. E. Spina, and D. Poletti, 
\href{https://doi.org/10.1103/PhysRevE.95.062202}{Phys. Rev. E, \textbf{95}, (2017)}; M. Lenz, S. W\"uster,  C. J. Vale, N. R. Heckenberg, H. Rubinsztein-Dunlop,  C. A. Holmes,  G. J. Milburn,  and M. J. Davis, 
\href{https://link.aps.org/doi/10.1103/PhysRevA.88.013635}{Phys. Rev. A, \textbf{88}, (2013)}; G. Lemos, R. Gomes, S. Walborn,  et al. 
\href{https://doi.org/10.1038/ncomms2214}{Nat Commun 3, 1211 (2012)}. 
\bibitem{Smolin 1986}
L. Smolin,
\href{https://doi.org/10.1016/0375-9601(86)90661-4}{Physics Letters A, \textbf{113} (1986)}.
\bibitem{Nelson 1966}
E. Nelson,
\href{https://link.aps.org/doi/10.1103/PhysRev.150.1079}{Phys. Rev. \textbf{150}, 1079 (1966)}.

\bibitem{Guerra 1983}
F. Guerra,  and L. M. Morato, 
\href{https://link.aps.org/doi/10.1103/PhysRevD.27.1774}{Phys. Rev. D \textbf{27}, 1774 (1983)}.

\bibitem{Pena 1983}
L. de la Pena and A. M. Cetto,
\href{https://doi.org/10.1063/1.523448}{Journal of Mathematical Physics \textbf{18}, 1612 (1977)}.
\bibitem{Santos 2006}
E. Santos,
\href{https://doi.org/10.1016/j.physleta.2005.11.039}{Physics Letters A \textbf{352}, 49 (2006)}. 
















\end{thebibliography}
\end{document}